\documentclass[twocolumn,showpacs,preprintnumbers,amsmath,amssymb,prl]{revtex4}

\usepackage{graphicx}
\usepackage{dcolumn}
\usepackage{bm}

\begin{document}

\title{Coherent oscillations in a superconducting tunable
flux qubit manipulated without microwaves}

\author{S. Poletto$^1$, F. Chiarello$^2$, M. G. Castellano$^2$,
J. Lisenfeld$^1$, A. Lukashenko$^1$, C. Cosmelli$^3$,
G. Torrioli$^2$, P. Carelli$^4$ and A. V. Ustinov$^1$}
\email{ustinov@physik.uni-karlsruhe.de}

\affiliation{$^1$ Physikalisches Institut, Universit{\"a}t
Karlsruhe (TH), D-76131 Karlsruhe, Germany\\
$^2$ Istituto di Fotonica e Nanotecnologie, CNR, 00156 Roma, Italy\\
$^3$ Dip. Fisica, Universit\`a di Roma La Sapienza, 00185 Roma, Italy\\
$^4$ Dip. Ingegneria Elettrica, Universit\`a dell'Aquila,
67040 Monteluco di Roio, Italy}

\date{\today}

\begin{abstract}
We experimentally demonstrate the coherent oscillations of a
tunable superconducting flux qubit by manipulating its energy
potential with a nanosecond-long pulse of magnetic flux. The
occupation probabilities of two persistent current states
oscillate at a frequency ranging from 6 GHz to 21 GHz, tunable
via the amplitude of the flux pulse. The demonstrated operation mode
allows to realize quantum gates which take less than 100 ps time and are
thus much faster compared to other superconducting qubits. An other
advantage of this type of qubit is its insensitivity to both thermal
and magnetic field fluctuations.
\end{abstract}


\maketitle

Superconducting qubits stand between the most promising systems
for the realization of quantum computation. Coherent quantum
evolution and manipulation have been demonstrated and extensively
studied for single \cite{Nakamura-99, Martinis-02, Mooij-03,
Vion-02, Koch-06} and coupled superconducting qubits
\cite{Berkley-02, Pashkin-03, Yamamoto-03, Mooij-07, Simmonds-07,
Chow-07}. In most cases, the state of superconducting qubits are
manipulated by means of  microwave pulses, with a technique
similar to the NMR manipulation of atoms. An alternative way to
manipulate qubits is based on modifying their energy potential
without applying any microwave signals \cite{Nakamura-99,Koch-06}.
The latter approach requires a much simpler experimental technique
and offers the possibility of using classical logic signals to
control a quantum processor \emph{in situ}, which is advantageous
for the large scale implementation of a quantum circuits.

In this Letter, we report the observation of tunable coherent
oscillations in a SQUID-based flux qubit. These oscillations are
obtained by manipulating the qubit with nanosecond-long pulses of
magnetic flux rather than microwaves. By this technique, we could
increase the oscillation frequency up to 21 GHz, which allows to
perform very fast logical quantum gates. Since the relevant quality
factor of a qubit is the number of gate operations which can be
performed during its coherence time, this result is of particular interest
towards the realization of a solid-state quantum computer.

\begin{figure}
\includegraphics[width=8.0cm]{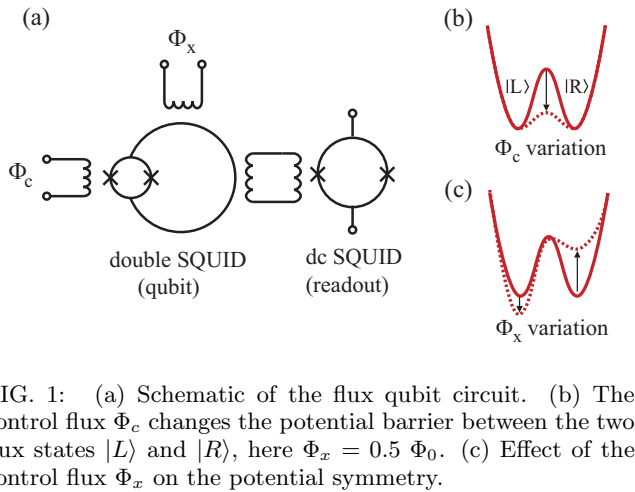}
\caption{ (a) Schematic of the flux qubit circuit. (b) The control
flux $\Phi _{c}$ changes the potential
barrier between the two flux states $|L\rangle$ and $|R\rangle$,
here $\Phi_x = 0.5\;\Phi_0$.
(c) Effect of the control flux $\Phi _{x}$ on the potential
symmetry. } \label{fig:1}
\end{figure}

The investigated circuit, shown in Fig.~\ref{fig:1}(a), is a
double SQUID consisting of a superconducting loop of inductance
$L=85$ pH, interrupted by a small dc SQUID of loop inductance
$l=6$ pH. This dc SQUID is operated as a single Josephson junction
(JJ) whose critical current is tunable by an external magnetic
field. Each of the two JJs embedded in the dc SQUID has a critical
current $I_{0} = 8 \mu$A and capacitance $C=0.4$ pF. The qubit is
manipulated by changing two magnetic fluxes $\Phi _{x}$ and
$\Phi_{c}$, applied to the large and small loops by means of two
coils of mutual inductance $M_{x} =2.6$ pH and $M_{c} =6.3$ pH,
respectively. The readout of the qubit flux is performed by
measuring the switching  current of an unshunted dc SQUID, which
is inductively coupled to the qubit \cite{Cosmelli-02}. The
circuit was manufactured by Hypres~\cite{hypres} using standard
Nb/AlO$_{x}$/Nb technology in a 100 A/cm$^2$ critical current
density process. The dielectric material used for junction
isolation is SiO$_2$. The whole circuit is designed
gradiometrically in order to reduce magnetic flux pick-up and
spurious flux couplings between the loops. The JJs have dimensions
of $3\times 3 \mu\mathrm{m}^2$ and the entire device occupied a
space of $230\times 430\mu \mathrm{m}^2$. All the measurements
have been performed at a sample temperature of 15 mK. The currents
generating the two fluxes $\Phi _{x}$ and $\Phi _{c}$ were
supplied via coaxial cables including 10 dB attenuators at the
1K-pot stage of a dilution refrigerator. To generate the flux
$\Phi _{c}$, a bias-tee at room temperature was used to combine
the outputs of a current source and a pulse generator. For biasing
and sensing the readout dc SQUID, we used superconducting wires
and metal powder filters \cite{Lukashenko-08} at the base
temperature, as well as attenuators and low-pass filters with a
cut-off frequency of 10 kHz at the 1K-pot stage. The chip holder
with the powder filters was surrounded by one superconducting and
two cryoperm shields.

Assuming identical junctions and negligible inductance of the
smaller loop ($l \ll L$), the system dynamics is equivalent to the
motion of a particle with the Hamiltonian
\[H=\frac{p^{2} }{2M} +\frac{\Phi _{b}^{2} }{L}
\left[\frac{1}{2} (\varphi -\varphi _{x} )^{2} - \beta (\varphi
_{c} )\cos \varphi \right],\] where $\varphi =\Phi /\Phi _{b}$ is
the spatial coordinate of the equivalent particle, $p$ is the
relative conjugate momentum, $M=C\Phi _{b} ^{2}$ is the effective
mass, $\varphi_{x} =\Phi _{x} /\Phi _{b}$ and $\varphi _{c} =\pi
\, \Phi _{c} /\Phi _{0}$ are the normalized flux controls, and
$\beta (\varphi _{c} )=(2I_{0} L/\Phi _{b})\cos \varphi _{c} $,
with $\Phi _{0} =h/\left(2e\right)$ and $\Phi _{b} =\Phi _{0}
/\left(2\pi \right)$. For $\beta <1$ the potential has a single
minimum, otherwise it consists of multiple wells. In the
particular case of $1 < \beta < 4.6$ and $\Phi _{x} =\Phi _{0}
/2$, the system potential is a symmetric double well shown in
Fig.~\ref{fig:1}(b). The two states $|L\rangle$ and $|R\rangle$,
which are respectively localized in the left and right potential
well, correspond to a persistent current circulating either
clockwise or counter-clockwise in the main SQUID loop. As it is
shown in Fig.~\ref{fig:1}(b), the external flux $\Phi _{c}$
controls the height of the barrier separating the minima, while a
variation of $\Phi _{x} $ changes the symmetry of the potential as
indicated in Fig.~\ref{fig:1}(c). In this work, we exploit both
the double well and the single well properties. The double well
potential shape is used for qubit initialization and readout. The
single well, or more exactly the two lowest energy states
$|0\rangle$ and $|1\rangle$ in this well, is used for the coherent
evolution of the qubit.

\begin{figure}
\includegraphics[width=8.0cm]{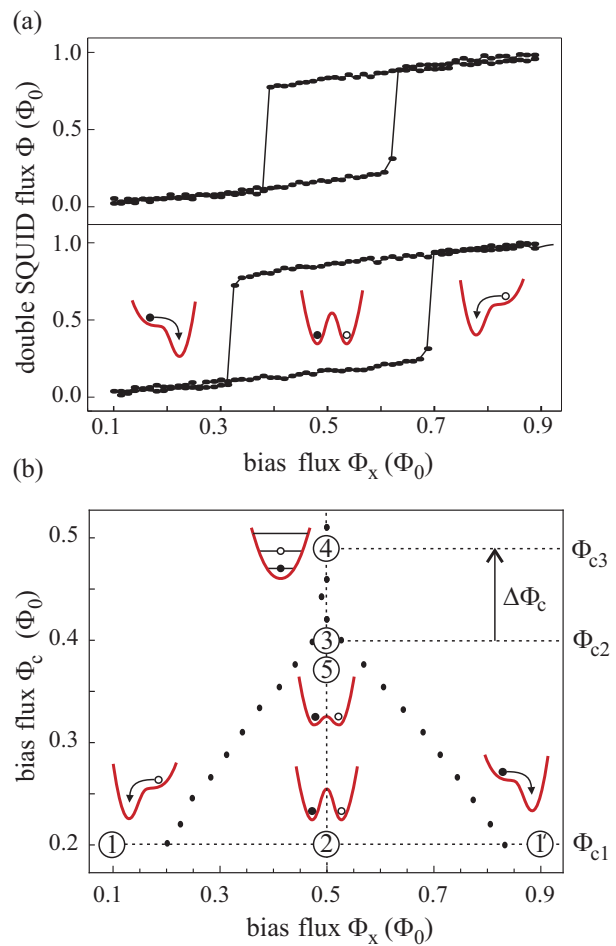}
\caption{(a) The measured double SQUID flux $\Phi$ in dependence of
$\Phi_x$, plotted for two different values of $\Phi_c$ and initial
preparation in either potential well. (b) Position of the
switching points (dots) in the $\Phi_c - \Phi_x$ parameter space.
Numbered tags indicate the working points for qubit manipulation
at which the qubit potential has a shape as indicated in the
insets. } \label{fig:2}
\end{figure}

We use a well established procedure \cite{Castellano-07} to
identify the regions where the system has a double well potential
in the $\Phi_c$ - $\Phi_x$ plane. The flux response $\Phi$ of the
qubit is measured as a function of $\Phi_x$ and $\Phi_c$ fluxes
and the switching points between different flux states are
detected. Figure \ref{fig:2}(a) shows $\Phi$ - $\Phi_x$
characteristics obtained for two $\Phi_c$ values using initial
state preparation in different wells. One can easily identify here
a bi-stability region in the vicinity of $\Phi_x \approx 0.5~
\Phi_0$ and a mono-stability region outside the hysteretic curve,
corresponding to the presence of a double and a single potential
well, respectively. At the border between these regions we find
abrupt switching between the two stable states. The positions
$\Phi_x$ of the switching points are plotted in
Fig.~\ref{fig:2}(b) for different $\Phi_c$ values. This diagram
allows to easily identify the combinations of parameters resulting
in a single- or double-well potential. A single-well region is
found for $\Phi_c\gtrsim 0.42$.

\begin{figure}
\includegraphics[width=8.0cm]{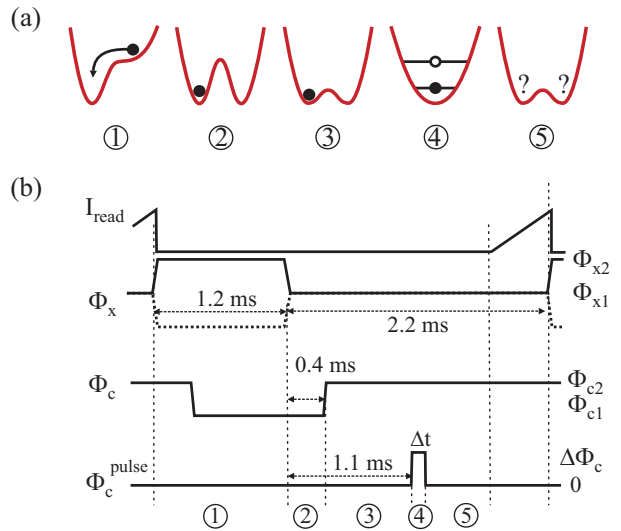}
\caption{(a) Variation of the potential shape during the
manipulation. (b) Time sequence of the readout dc SQUID current
(topmost line) and flux bias values (bottom lines). }
\label{fig:3}
\end{figure}

The measurement process that we used to observe coherent
oscillations consists of several steps shown in
Fig.~\ref{fig:3}(a). Each step is realized by applying a
combination of magnetic fluxes $\Phi_x$ and $\Phi_c$ as indicated
by numbers in Fig.~\ref{fig:2}(b). The first step in our
measurement is the initialization of the system in a defined flux
state (1).  Starting from a double well at $\Phi _{x} \cong \Phi
_{0} /2$ with high barrier, the potential is tilted by changing
$\Phi_x$  until it has only a single minimum (left or right,
depending on the amplitude and polarity of the applied flux
pulse). This potential shape is maintained long enough to ensure
the relaxation to the ground state. Afterwards the potential is
tuned back to the initial double-well state (2). The high barrier
prevents any tunneling and the qubit is thus initialized in the
chosen potential well. Next, the barrier height is lowered to
an intermediate level (3) that preserves the initial state and
allows to use just a small-amplitude $\Phi_c$ flux pulse for the
subsequent manipulation. The following $\Phi_c$-pulse transforms
the potential into a single well (4). The $\Phi_c$-pulse duration
$\Delta t$ is in the nanosecond range. The relative phase of the
ground and the first excited states evolves depending on the energy
difference between them. Once $\Phi_c$-pulse is over, the double
well is restored and the system is measured in the basis $\{
|L\rangle ,|R\rangle \} $ (5). The readout of the qubit flux state
is done by applying a bias current ramp to the dc SQUID and
recording its switching current to the voltage state.

The pattern that realizes the above described manipulation is
reported in Fig.~\ref{fig:3}(b). The flux  $\Phi_x$ is switched
between two values: $\Phi_{x2}$ is used to create a strongly
asymmetric potential for qubit initialization in the left or right
well, and $\Phi_{x1}$ equal to $\Phi_0/2$ (or very close to it)
transforms the potential into a symmetric (or nearly
symmetric) double well. The flux $\Phi_c$ is changed between three
different values: $\Phi_{c1}$ and $\Phi_{c2}$ define,
respectively, high and intermediate amplitudes of the barrier
between the two minima, while at $\Phi_{c3} = \Phi_{c2} + \Delta
\Phi_c$ the barrier is removed completely and the potential turns
into a single well. The amplitude $\Delta \Phi_c$ of the pulse is
varied allowing for single wells of different curvature at the
bottom. The nominal rise and fall times of this pulse are
$t_{r/f}=0.6$ ns.

\begin{figure}
\includegraphics[width=8.0cm]{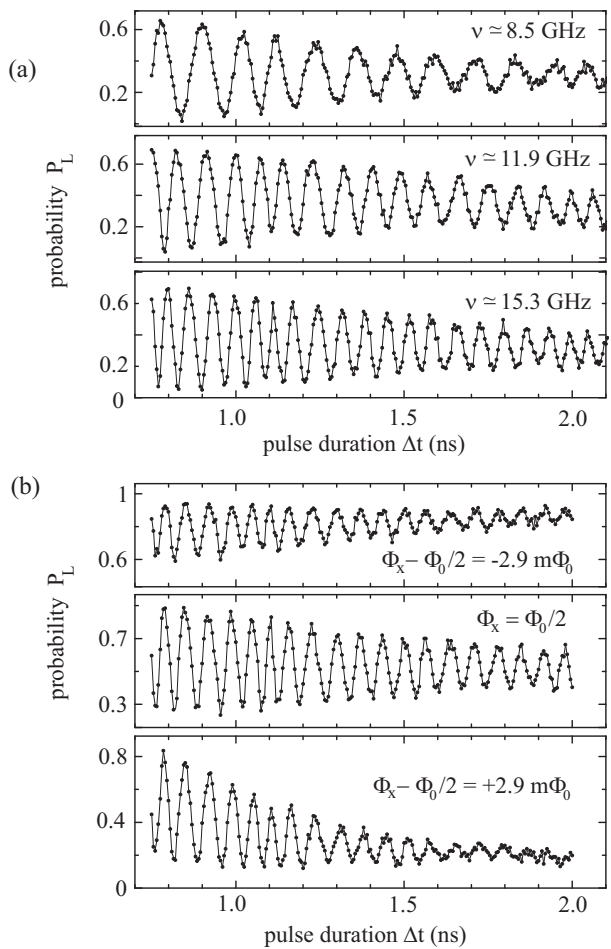}
\caption{Probability to measure the state $\left| L \right\rangle$
in dependence of the pulse duration $\Delta t$ for the qubit initially
prepared in the $\left| L \right\rangle$ state, and for
(a) different pulse amplitudes $\Delta \Phi_c$, resulting
in the indicated oscillation frequency, and
(b) for different potential symmetry by detuning $\Phi_x$
from $\Phi_0/2$ by the indicated amount.}
\label{fig:4}
\end{figure}

The flux pattern is repeated for $10^2 - 10^4$ times in order to
evaluate the probability $P_{L} =\left | \left \langle L | \Psi
_\mathrm{final} \right \rangle \right |^{2}$ of occupation of the
left state at the end of the manipulation. By changing the
duration $\Delta t$ of the manipulation pulse $\Phi_c$, we observed
coherent oscillations between the occupations of the states
$|L\rangle$ and $|R\rangle$ shown in Fig.~\ref{fig:4}(a). The
oscillation frequency could be tuned between 6 and 21 GHz by
changing the pulse amplitude $\Delta \Phi_c$.
These oscillations persist when the
potential is made slightly asymmetric by varying the value
$\Phi_{x1}$. As it is shown in Fig.~\ref{fig:4}(b), detuning from
the symmetric potential by up to $\pm 2.9 \mathrm{m}\Phi_0$ only
slightly changes the amplitude and symmetry of the oscillations.
When the qubit was initially prepared in $|R\rangle$ state instead
of $|L\rangle$ state we observed similar oscillations.

To understand the physical process behind the observed
oscillations, let us discuss in detail what happens during the
manipulation. Suppose the system is initially prepared in the left
state $\left| L \right\rangle$ of a perfectly symmetric double
well potential. During the $\Phi_c$ pulse, the potential has only
one central minimum and can be approximated by a harmonic
oscillator potential with frequency $\omega _{0} \left(\Phi_{c3}
\right)\approx 1/\sqrt{2LC}\, \sqrt{1-\beta \left(\Phi_{c3}
\right)}$. The pulse transforms the initially prepared left state
(that is a symmetric superposition of the two lowest energy
eigenstates of the double well potential $\left| \tilde 0
\right\rangle$ and  $\left| \tilde 1 \right\rangle$, i.e. $\left|
L \right\rangle=(\left| \tilde 0\right\rangle+\left| \tilde 1
\right\rangle)/\sqrt{2}$ ) into the superposition of the two
lowest energy eigenstates $\left| 0\right\rangle$ and $\left| 1
\right\rangle$ of the single-well potential. To achieve that, the
pulse rise time needs to be shorter than the relaxation time but,
at the same time, long enough to avoid population of upper energy
levels. During the plateau of the $\Delta \Phi_c$ pulse, the
relative phase $\theta $ between the states $\left| 0
\right\rangle$ and $\left| 1 \right\rangle$ evolves with time at
the Larmor frequency given by $\omega _{0} =\left(E_{1} -E_{0}
\right)/\hbar$. At the end of the pulse the accumulated relative
phase becomes $\theta =\omega _{0} \Delta t$. Turning the system
back into the double-well maps the phase to the two flux states
$\left| L \right\rangle$ and  $\left| R \right\rangle$. The final
state after the flux pulse $\Phi_c$ is $\left |\Psi_\mathrm{final}
\right\rangle =\cos ( \theta /2)\, \left| L \right\rangle + i \sin
(\theta /2)\left | R \right \rangle$. Note that in the more
realistic case of non perfectly symmetric double-well potential,
the initial left state is no more a symmetric superposition, but
tends more to either $\left| \tilde 0 \right\rangle$ or $\left|
\tilde 1 \right\rangle$ due to the potential unbalancing. However,
a pulse with a short rise time induces a non-adiabatic transition
that populates mainly the two lowest energy eigenstates $\left| 0
\right\rangle$ and $\left| 1 \right\rangle$ in the single well
potential. This condition can be met in a narrow region of the
flux bias plane called ``portal'' \cite{Koch-05}. This
non-adiabatic transition also leads to the phase evolution process
described above.

In order to verify above interpretation, we numerically solved the
time-dependent Schr\"odinger equation for this system. The
simulation showed that with our experimental parameters the
transition between the first two levels occurs as described, while
the occupation of upper levels remains below few percents.

\begin{figure}
\includegraphics[width=7.0cm]{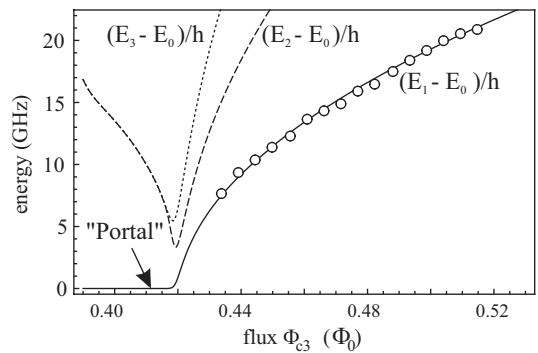}
\caption{Calculated energy spacing of the first (solid line),
second (dashed line) and third (dotted line) energy levels with
respect to the ground state in the single well potential, plotted
vs. the control flux amplitude $\Phi_{c3}$. Circles are the
experimentally observed oscillation frequencies for the
corresponding pulse amplitudes.} \label{fig:5}
\end{figure}

The oscillation frequency $\omega_{0}$ depends on the amplitude of
the manipulation pulse $\Delta \Phi_c$ since it determines the
shape of the single well potential and the energy level spacing
$E_{1}-E_{0}$. A pulse of larger amplitude $\Delta \Phi_{c}$
generates a deeper well having a larger level spacing, which leads
to a larger oscillation frequency as shown in Fig.~\ref{fig:4}(a).
In Fig.~\ref{fig:5}, we plot the energy spacing between the ground
state and the three excited states (indicated as $\left(E_{k}
-E_{0} \right)/h$ with k=1,2,3) versus the flux $\Phi _{c3} =\Phi
_{c2} +\Delta \Phi_{c} $ obtained from a numerical simulation of
our system using the experimental parameters. In the same figure,
we plot the measured oscillation frequencies for different values
of $\Phi_{c}$ (open circles). Excellent agreement between
simulation (solid line) and data strongly supports our
interpretation. The fact that a small asymmetry in the potential
does not change the oscillation frequency, as shown in
Fig.~\ref{fig:4}(b), is consistent with the interpretation as the
energy spacing $E_{1} - E_{0} $ is only weakly affected by small
variations of $\Phi _{x}$. This provides protection against noise
in the controlling flux $\Phi_x$.

The measured oscillation decay time of about 2 ns is of the same
order as the Rabi oscillations decay and the energy relaxation
time $T_1$ which we measured by using standard microwave
$\pi$-pulse manipulation on the same device. It is also comparable
to the coherence time obtained on similar devices fabricated using
the same technology \cite{Lisenfeld-07}, suggesting that coherence
is not limited by the manipulation procedure reported in this
paper. We believe that the decay time of the reported
high-frequency coherent oscillations can be increased by two
orders of magnitude by reducing the area of the JJs and using an
appropriate dielectric instead of SiO$_{2}$ as insulating material
in the junction fabrication \cite{Martinis-05}.

We note that a similar qubit manipulation procedure has been
reported by Koch et al.~\cite{Koch-06}, demonstrating Larmor
oscillations in a flux qubit coupled to a harmonic oscillator. It
should be emphasized that, in our case, the oscillator is not
required, which simplifies the realization of the qubit circuit.
Moreover, in contrast to Ref.~\cite{Koch-06}, our approach
provides a \emph{wide range tunability} of the frequency of
coherent oscillations. This allows for a variety of quantum gates
by manipulating both duration and amplitude of the flux
pulses, and moreover is important concerning the realization of a controllable
coupling between qubits or quantum busses such as resonant cavities.

In conclusion, we presented the coherent manipulation of a flux
qubit without using microwaves. The reported approach seems
particularly promising for the realization of circuits with many
qubits, and it appears to be well suited for integration with RSFQ
control electronics~\cite{Feldman-01etc}. The benefits of the
reported system are the possibility of \emph{in situ} tuning the
frequency of oscillations and their insensitivity to small changes
in the potential symmetry. The high frequency of oscillations
allows for very fast qubit gate operations, and the large energy
gap between the qubit states during coherent evolution protects the
system from thermal activation to upper energy states. Moreover,
the oscillation frequency depends only weakly on the control pulse
amplitude $\Delta\Phi_c$, in contrast to the exponential
sensitivity of the oscillations in a double well potential
\cite{Chiarello-07}, which makes the qubit manipulation more
reliable.

This work was partially supported by the Deutsche
Forschungsgemeinschaft (DFG), the CNR RSTL program
and the EU projects RSFQubit and EuroSQIP.


\begin{thebibliography}{0}

\bibitem{Nakamura-99} Y. Nakamura, Y. A. Pashkin, and J. S. Tsai,
Nature (London) \textbf{398}, 786 (1999).

\bibitem{Martinis-02} J.M. Martinis \textit{et al.}, Phys. Rev. Lett. \textbf{89}, 117901 (2002).

\bibitem{Mooij-03} I. Chiorescu \textit{et al.}, Science \textbf{299}, 1869 (2003).

\bibitem{Vion-02} D. Vion \textit{et al.}, Science \textbf{296}, 886 (2002).

\bibitem{Koch-06} R. H. Koch \textit{et al.}, Phys. Rev. Lett. \textbf{96}, 127001 (2006).

\bibitem{Berkley-02} A. J. Berkley \textit{et al.}, Science \textbf{300}, 1548 (2003).

\bibitem{Pashkin-03} Y. A. Pashkin \textit{et al.}, Nature (London) \textbf{421} , 823 (2003).

\bibitem{Yamamoto-03} T. Yamamoto \textit{et al.}, Nature (London) \textbf{425}, 941 (2003).

\bibitem{Mooij-07} J.H. Plantenberg \textit{et al.}, Nature (London) \textbf{447}, 836 (2007).

\bibitem{Simmonds-07} M. A. Sillanp\"a\"a, J. I. Park, and R. W. Simmonds,
Nature (London) \textbf{449}, 438 (2007).

\bibitem{Chow-07} J. Majer \textit{et al.}, Nature (London) \textbf{449}, 443 (2007).

\bibitem{Cosmelli-02} C. Cosmelli \textit{et al.}, Physica C \textbf{372}, 213 (2002).

\bibitem{hypres} Hypres Inc., Elmsford, N.Y., USA.

\bibitem{Lukashenko-08} A. Lukashenko and A. V. Ustinov, Rev. Sci. Instr.
\textbf{79}, 014701 (2008).

\bibitem{Castellano-07} M. G. Castellano \textit{et al.}, Phys. Rev. Lett. \textbf{98}, 177002 (2007).

\bibitem{Koch-05} R. H. Koch \textit{et al.}, Phys. Rev. B \textbf{72}, 092512 (2005).

\bibitem{Lisenfeld-07} J. Lisenfeld \textit{et al.}, Phys. Rev. Lett. \textbf{99}, 170504 (2007).

\bibitem{Martinis-05} J. M. Martinis \textit{et al.}, Phys. Rev. Lett. \textbf{95}, 210503 (2005).

\bibitem{Feldman-01etc} M. J. Feldman and M. F. Bocko, Physica C \textbf{350}, 171
(2001); T. A. Ohki, M. Wulf, and M. J. Feldman, IEEE Trans. Appl.
Supercond. \textbf{17}, 154 (2007).

\bibitem{Chiarello-07} F. Chiarello, Eur. Phys. J. B \textbf{55}, 7 (2007).

\end{thebibliography}
\end{document}